\documentclass[twocolumn,prb,showpacs,amsmath,amssymb,superscriptaddress,nofootinbib,floatfix]{revtex4-1}
\usepackage{graphicx}
\usepackage{dcolumn}
\usepackage{bm}
\usepackage{color}
\usepackage{soul}

\newcommand{\be}{\begin{equation}}
\newcommand{\e}{\end{equation}}
\newcommand{\beml}{\begin{subequations}}
\newcommand{\eml}{\end{subequations}}
\newcommand{\beq}{\begin{eqnarray}}
\newcommand{\eq}{\end{eqnarray}}
\newcommand{\ba}{\begin{array}}
\newcommand{\ea}{\end{array}}
\newcommand{\bpm}{\begin{pmatrix}}
\newcommand{\epm}{\end{pmatrix}}
\newcommand{\bc}{\begin{cases}}
\newcommand{\ec}{\end{cases}}
\newcommand{\lt}{\left}
\newcommand{\rt}{\right}

\newcommand{\ep}{\varepsilon}

\begin{document}

\title{Magnetotransport in single layer graphene in a large parallel magnetic field}

\author{F.~Chiappini}
\email[]{f.chiappini@science.ru.nl}
\affiliation{Radboud University, High Field Magnet Laboratory (HFML-EMFL), NL-6525 ED Nijmegen, The Netherlands}
\affiliation{Radboud University, Institute for Molecules and Materials, NL-6525 AJ Nijmegen, The Netherlands}
\author{S.~Wiedmann}
\affiliation{Radboud University, High Field Magnet Laboratory (HFML-EMFL), NL-6525 ED Nijmegen, The Netherlands}
\affiliation{Radboud University, Institute for Molecules and Materials, NL-6525 AJ Nijmegen, The Netherlands}
\author{M.~Titov}
\affiliation{Radboud University, Institute for Molecules and Materials, NL-6525 AJ Nijmegen, The Netherlands}
\author{A.~K.~Geim}
\affiliation{School of Physics and Astronomy, University of Manchester, Oxford Road, Manchester, M13 9PL, UK}
\author{R.~V.~Gorbachev}
\affiliation{School of Physics and Astronomy, University of Manchester, Oxford Road, Manchester, M13 9PL, UK}
\author{E.~Khestanova}
\affiliation{School of Physics and Astronomy, University of Manchester, Oxford Road, Manchester, M13 9PL, UK}
\author{A.~Mishchenko}
\affiliation{School of Physics and Astronomy, University of Manchester, Oxford Road, Manchester, M13 9PL, UK}
\author{K.~S.~Novoselov}
\affiliation{School of Physics and Astronomy, University of Manchester, Oxford Road, Manchester, M13 9PL, UK}
\author{J.\,C.~Maan}
\affiliation{Radboud University, High Field Magnet Laboratory (HFML-EMFL), NL-6525 ED Nijmegen, The Netherlands}
\affiliation{Radboud University, Institute for Molecules and Materials, NL-6525 AJ Nijmegen, The Netherlands}
%\affiliation{High Field Magnet Laboratory (HFML - EMFL), Radboud University, Toernooiveld 7, 6525 ED NIJMEGEN, The Netherlands.}
\author{U.~Zeitler}
\email[]{u.zeitler@science.ru.nl}
\affiliation{Radboud University, High Field Magnet Laboratory (HFML-EMFL), NL-6525 ED Nijmegen, The Netherlands}
\affiliation{Radboud University, Institute for Molecules and Materials, NL-6525 AJ Nijmegen, The Netherlands}

\date{\today}

\begin{abstract}
Graphene on hexagonal boron-nitride (h-BN) is an atomically flat conducting system that is ideally suited for probing the effect of Zeeman splitting on electron transport. We demonstrate by magneto-transport measurements that a parallel magnetic field up to 30 Tesla does not affect the transport properties of graphene on h-BN even at charge neutrality where such an effect is expected to be maximal. The only magnetoresistance detected at low carrier concentrations is shown to be associated with a small perpendicular component of the field which cannot be fully eliminated in the experiment. Despite the high mobility of charge carries at low temperatures, we argue that the effects of Zeeman splitting are fully masked by electrostatic potential fluctuations at charge neutrality.
\end{abstract}


\maketitle

\section{Introduction}
A magnetic field applied in the plane of an ideally flat two-dimensional (2D) conductor couples to the spin degree of freedom of charge carriers rather than to their orbital motion. In such a setup, the orbital effects such as Hall and Nernst are suppressed and the spin-polarization effects become the leading phenomena.
This idea has been intensely exploited in semiconductor heterostructures to study the effects of electron-electron interactions and disorder on spin polarisation and spin-resolved density of states in two-dimensional electron gases (2DEGs).\cite{TutucPRL2002,OkamotoPRL1999}

For some 2DEGs, the characteristic width of the confinement potential is, however, comparable to the magnetic length $\ell_B= \sqrt{\hbar/eB}$ even for fields of the order of a few Tesla. For larger fields the energy bands and, consequently, the effective mass and the \textit{g}-factor of electrons become sensitive to the value of the in-plane magnetic field $B_\parallel$ \cite{DasSarmaPRL2000,ZhouPRL2010} and the interplay between spin and orbital effects influences the transport properties of the system.\cite{PiotPRB2009, TutucPRB2003}
%These additional effects strongly complicate the interpretation of experimental data.\cite{PiotPRB2009, TutucPRB2003}

In contrast to semiconductor-based 2DEGs, a complete decoupling of the orbital and spin effects can be achieved in graphene. %Indeed, the intrinsic spin-orbit coupling is negligible for temperatures $> 10$\,mK.\cite{MinPRB2006}
Since graphene is only one atom thick, the orbital motion of the electrons is not affected by $B_\parallel$ up to the fields of the order of $10^3$\,T. In actual devices, however, graphene adapts to the conformation of the underlying substrate.\cite{LuiNature2009} Common substrates such as SiO$_2$ induce corrugations (ripples) to graphene plane that convert a nominal in-plane field into a randomly oriented one, depending on the curvature of the surface.  Experimental works on SiO$_2$ supported graphene showed that the external $B_\parallel$ couples to the orbital motion of carriers via the high corrugations leading to a magnetoresistance which depends on the topography of the device \cite{LundebergPRL2010, WakabayashiJoP2011}.

Nevertheless, an atomically flat conducting system can be achieved by placing graphene on hexagonal boron-nitride (h-BN) which significantly increases its mobility.\cite{DeanNatNanotech2010, XueNatMat2011} Graphene sandwiched between two atomically flat h-BN surfaces gives rise to an ultimately sharp potential well with a characteristic width of one atom\cite{HaighNatMat2012}, representing an ideal playground to probe the effects of an in-plane magnetic field on the electron transport of a truly 2D system.

An in-plane magnetic field modifies the density of states only due to the Zeeman splitting $E_Z=g\mu_BB$, where $g=2$ is the electron g-factor and $\mu_B$ is the Bohr magneton, leading to a value $E_Z\approx 3.5$\,meV for $B= 30$\,T. In addition, the splitting of spin sub-bands changes the {density of states in graphene at charge neutrality from zero ta a non-zero value} leading to a non-zero quasiparticle density $n_\textrm{Q}$. Therefore, a strong in-plane magnetic field is expected to affect magneto-transport properties of graphene only in the limit of low charge carrier density, $n<n_\textrm{Q}$, and low temperature, $T< E_Z$.\cite{HwangPRB2009}

In this work, we investigate the resistivity of high-quality h-BN supported graphene  in the presence of a large in plane magnetic field. We do not observe any change of resistivity induced by  $B_\parallel$ neither at charge neutrality nor for large doping at 1.4\,K and for $B_\parallel$ as large as $30$\,T. Despite the high mobility of charge carriers in the sample  $\mu\approx 50000$\,cm$^{2}$V$^{-1}$s$^{-1}$, the electrostatic potential fluctuations around the charge neutrality point (CNP) are sufficiently strong to average out possible effects of Zeeman splitting.

\section{Experimental details}

Our sample is a Hall-bar shaped graphene device with an aspect ratio $L/W=2$ (the distance between contacts $L\approx 3$\,$\mu$m and the width $W\approx 1.5$\,$\mu$m). The graphene flake, sandwiched between two thin layers of h-BN, is connected to Ti/Au contacts. The system is placed on top of a doped Si/SiO$_{2}$ wafer, which acts as a back gate. Low temperature ($T=1.4$\,K) transport measurements were performed using a low frequency lock-in technique with a $10$\,nA excitation. The longitudinal $\rho_{xx}$ and Hall $\rho_{xy}$ resistivities were measured as a function of the back gate voltage $V_\textrm{G}$ and the external magnetic field $B$ that varies up to 30\,T.

%%%%%%%%%%%%%%%%%%%%%%%%%%%%
%%%% fig:fig1
%%%%%%%%%%%%%%%%%%%%%%%%%%%%
\begin{figure}[t]
\centerline{\includegraphics[width=0.48\textwidth]{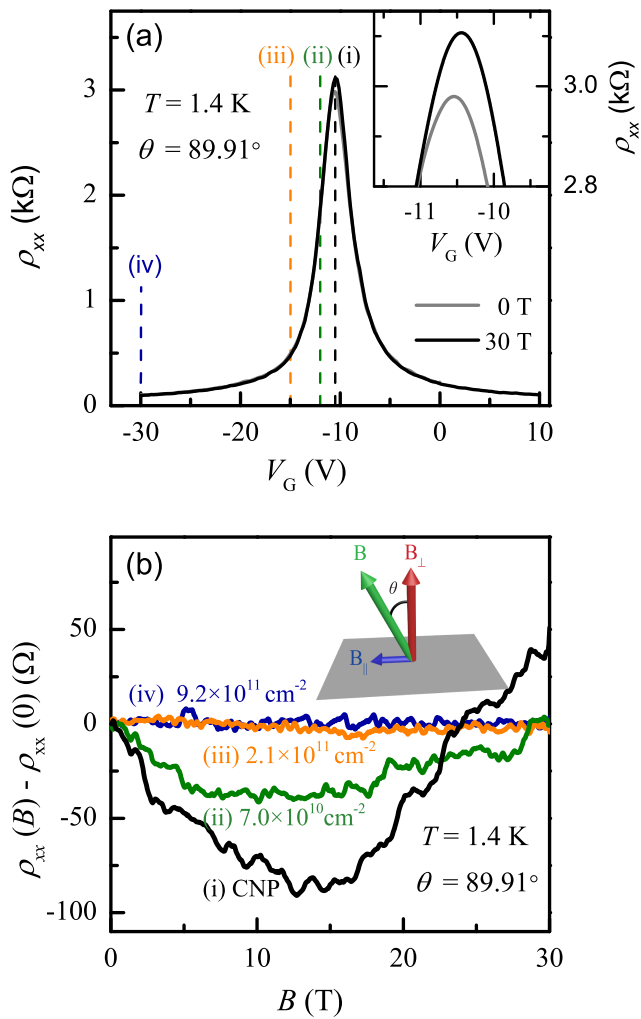}}
\caption{Panel (a): Resistivity $\rho_{xx}$ versus gate voltage $V_\textrm{G}$ for $B= 0$ (grey line) and $B= 30$\,T (black solid line). Inset: Resistivity $\rho_{xx}$ versus $V_\textrm{G}$ in the vicinity of the CNP. \newline
Panel (b): Magnetoresistance as a function of $B$ in the best parallel-field configuration, $\theta=89.91^\circ \pm 0.01^\circ$. Different lines are for different concentrations of charge carriers: corresponding gate voltages are indicated with the dashed lines in panel (a). Inset: Configuration of magnetic field orientation with respect to the graphene plane (shown in grey) and definition of tilt angle $\theta$. $\theta=90^\circ$ corresponds to a purely in-plane field.
}
\vspace{0em}
\label{fig:fig1}
\end{figure}
 %%%%%%%%%%%%%%%%%%%%%%%%%%%%%

The charge neutrality point is associated with the maximum of $\rho_{xx}$ at $V_\textrm{G}=-10.5$\,V (see Fig.~\ref{fig:fig1}a){and its shift with magnetic field is negligible, allowing measuring the resistance of the CNP with a relative accuracy of better than 0.3\% by sweeping the field at a constant gate voltage $V_\textrm{G}=-10.5$\,V.}
The charge carrier concentration $n$ is assumed to be proportional to the gate voltage $n= -\alpha\,(V_\textrm{G}-V_\textrm{CNP})$, where the proportionality coefficient is set by $\alpha=4.7\times10^{14}$\,m$^{-2}$V$^{-1}$. The value of $\alpha$ is obtained experimentally from the dependence of Shubnikov-de-Haas oscillations on $V_\textrm{G}$ for a given perpendicular component of the field $B_\perp$.

A linear fit of the conductivity at large $n$, $\sigma_{xx}=e\mu |n|$, gives rise to an estimate of the hole mobility in the system $\mu \approx 50000$\,cm$^{2}$V$^{-1}$s$^{-1}$. In what follows we focus mostly on the hole-doped region $V_\textrm{G}< V_\textrm{CNP}$ since the hole mobility in the sample appears to be higher than the electron one. The high quality of our sample is testified by the observation of the fully developed integer quantum Hall effect at $B_\perp=2.5$\,T and the observation of the lifting of the spin degeneracy of the Landau levels at $B_\perp=10$\,T.\cite{ChiappiniPRB2015}

The sample is mounted on a rotating stage with a single axis rotator that allows \textit{in situ} rotation at low temperature. We define $\theta$ as the angle between the direction of external magnetic field and the normal to the graphene plane as shown in the inset of Fig.~\ref{fig:fig1}b. For $\theta=90^\circ$ the field is entirely in plane, $B_\perp=B\cos\theta=0$.

The angle $\theta$ is estimated from the measurement of Hall resistivity by using the expression $\rho_{xy}=B\cos\theta(en)^{-1}$ which holds for sufficiently large $n$ in the single-component classical Hall regime. With our experimental setup we achieve $\theta=90^\circ$ within less than $0.1^\circ$, which corresponds to $B_\parallel \approx B$ and $B_\perp \lesssim 50$\,mT at the maximal applied field $B=30$\,T.

\section{Magnetotransport in a parallel magnetic field}

Fig.~\ref{fig:fig1}a shows the resistivity $\rho_{xx}$ as a function of the gate voltage $V_\textrm{G}$ for $\theta = 89.91^\circ$. {This was the closest} experimentally achievable angle to the parallel field configuration in our tilted-field setup. Note that this relative misalignment of less than $10^{-3}$ , corresponds to lateral displacement of the sample, mounted on a $\approx 1$~m long probe, of less than 1 mm.
The grey curve represents the signal in the absence of the field while the black curve corresponds to the external field $B =30$\,T.  Away from the CNP the two traces are indistinguishable. A small increase in resistivity is observed in the region around the CNP (see inset of Fig.\ref{fig:fig1}(a)) at maximum field.

To better illustrate the response of $\rho_{xx}$ to $B$, we plot in Fig.~\ref{fig:fig1}b  the magnetoresistance (defined as $\rho_{xx}(B)-\rho_{xx}(0)$) as a function of the magnetic field in the best parallel-field configuration for specific gate voltage values indicated by the dashed lines in Fig.~\ref{fig:fig1}a, corresponding to the CNP $n=0$ (black line), $n=7 \times 10^{10}$\,cm$^{-2}$ (green line), $n=2.1 \times 10^{11}$\,cm$^{-2}$ (orange line) and $n=9.2 \times 10^{11}$\,cm$^{-2}$ (blue line).

At high $n$ (blue and orange curves) the resistivity is not sensitive to $B\approx B_\parallel$ while a dependence $\rho_{xx}(B)$ is seen in a vicinity of the CNP (green and black lines).

One can clearly see  that the observed magnetoresistance is maximized at the CNP. It reaches a maximal negative value for $B\approx 15$\,T and it increases for larger fields. Eventually it changes sign at $B\approx 25$\,T. A similar non-monotonic behaviour can also be seen for $n= 7\times 10^{10}$ cm$^{-2}$ (green line), though $\rho_{xx}$ reaches its zero field value at $B\approx 30$\,T. {The angles were calibrated by measuring the Hall voltage at a large negative gate voltage $V_G = - 30$~V corresponding to a hole concentration $n = 1\times 10^{12}$~cm$^{-2}$, see Fig.~2b.}

However, this seemingly non-trivial magnetoresistance is induced entirely by the remaining perpendicular component of the field $B_\perp$, which cannot be ignored in the vicinity of the CNP.

In order to prove that the observed changes of $\rho_{xx}$ are indeed related to $B_\perp$, we measure the magnetoresistance for slightly different tilt angles around $\theta=90^{\circ}$.

\begin{figure}[t]
\centerline{\includegraphics[width=0.48\textwidth]{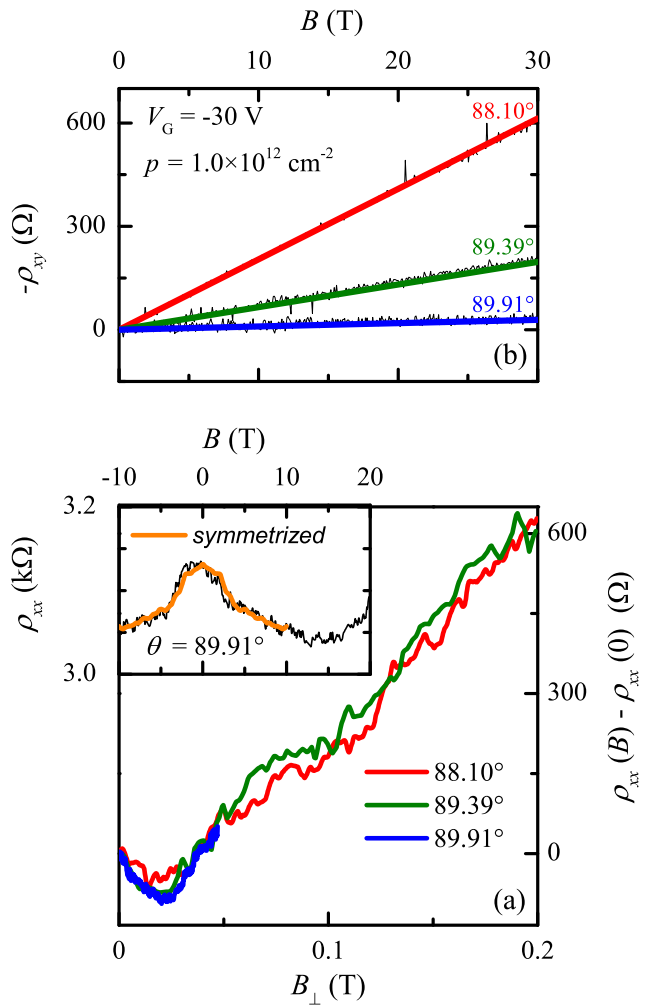}}
\caption{{Panel (a): }Magnetoresistance at the CNP as a function of $B_\perp$ at $1.4$\,K for three different angles $\theta$. {The inset illustrates the symmetry of $\rho_{xx}$ measured at $\theta = 89.91^\circ$ with the black curve the original data between -10~T an 20~T and the orange line the symmetrized curve between -10~T and 10~T.}\newline
{Panel (b):  Hall resistivity of the sample at a large hole concentration ($n = 1 \times 10^{12}$~cm$^{-2}$) measured as a function of total magnetic field for the same angles .
The solid lines represent the expected behavior $\rho_{xy} = - B \cdot \cos \theta / (ne)$ at the angles used.}
}
\vspace{0em}
\label{fig:fig2}
\end{figure}

In Fig.~\ref{fig:fig2}a we then plot the $\rho_{xx}$ data for three different angles: $\theta = 83.39^\circ$, $\theta = 88.1^\circ$ and $\theta=89.91^\circ$. The curves fall on top of each other when plotted with respect to $B_\perp = B \cos\theta$. These experimental results suggest that the magnetoresistance observed in the vicinity of the CNP for $\theta\approx 90^{\circ}$ is entirely due to the perpendicular component of the field and that $B_\parallel$ does not produce any sizeable effect in the resistivity of our device. {We have also checked that no contribution of $\rho_{xy}$ is superimposed onto $\rho_{xx}$ by measuring it for both field orientations and by symmetrizing the $\rho_{xx}$-data, see inset in Fig.~2a. Within experimental accuracy we find that $\rho_{xx}$ is even in magnetic field, i.e.~we can safely neglect any odd contributions from $\rho_{xy}$ onto it.}

%%%%%%%%%%%%%%%%%%%%%%%%%%%%
%%%% fig:fig2
%%%%%%%%%%%%%%%%%%%%%%%%%%%%
%%%%%%%%%%%%%%%%%%%%%%%%%%%%%

\section{Discussion}

The observed dependence $\rho_{xx}(B_\perp)$ at the CNP has already been addressed in numerous references and can be explained as follows:
The initial decrease in resistance is compatible with the suppression of weak localization\cite{OstorovskyPRB2006} due to external magnetic field. This phenomenon can be expected at such a low temperature and small $B_\perp$.\cite{TikhonenkoPRL2008}. For larger values of  $B_\perp$ the positive magnetoresistance can be associated to classical effects such as two-liquid transport (see e.\,g.\,Refs.~\onlinecite{WiedmannPRB2011}, \onlinecite{AlekseevPRL2015} and references therein) and a semiclassical linear magnetoresistance arising from concentration fluctuations.\cite{ChangSSC1985, Benno1997}. When moving away form the CNP all these effects rapidly decrease which is indeed observed experimentally in a strong suppression of  the observed magnetoresistance, see Fig.~\ref{fig:fig1}(b).

Let us now discuss the experimental results from the point of view of a simple Drude theory which does not take into account localisation phenomena.\cite{Weiss1954}  Assuming equal mobilities of electron- and hole-like quasiparticles, one obtains the resistivity tensor
\begin{equation}
\label{rho}
\rho_{xx}= \frac{n_\textrm{Q}}{e\mu}\frac{1+\mu^2 B_\perp^2}{n_\textrm{Q}^2+n^2 \mu^2 B_\perp^2},\quad
\rho_{xy}=\frac{n}{n_\textrm{Q}}\mu B_\perp \rho_{xx},
\end{equation}
which depends on two densities: the charge carrier density $n=n^h_{+}+n^h_{-} - n^e_{+}-n^e_{-}$ and the quasiparticle density  $n_\textrm{Q}=n^e_{+}+n^e_{-} + n^h_{+}+n^h_{-}$. Here, the electron and hole  densities, $n^e_{\sigma}$ and $n^h_{\sigma}$ correspondingly, are defined for different spin species $\sigma=\pm$  as
\be
\label{dens}
n^{e,h}_{\sigma} =\int_0^\infty \!\!\!\nu(\ep)\, f^{e,h}_\sigma(\ep)\,d\ep,
\e
where $f^e_\sigma(\ep)= \lt[1+\exp\lt[(\ep-\sigma E_Z/2-\mu_c)/T\rt]\rt]^{-1}$ is the electron Fermi distribution function, $f^h_\sigma(\ep)= 1-f^e_\sigma(-\ep)$, $\mu_c$ is the chemical potential, and $\nu(\ep)=\nu(-\ep)$ is density of states per spin which is taken to be symmetric with respect to the Dirac point. For ideally clean graphene in zero field  $\nu(\ep)=|\ep|/\pi\hbar^2v^2$.

At large doping ($n=\pm  n_\textrm{Q}$), one finds $\rho_{xx} = 1/e \mu |n|$, which lacks an explicit dependence on $B_\perp$.  At the CNP ($n=0$), one finds $\rho_{xx} = (e \mu n_\textrm{Q})^{-1} (1+\mu^2 B_\perp^2)$ which manifestly increases with $B_\perp$. The quadratic dependence on $B_\perp$ in the homogeneous Drude model is transformed into a linear one (which is clearly seen in Fig.~\ref{fig:fig2}) due to the boundary effects or large-scale electrostatic potentials variations.\cite{AlekseevPRL2015,VasilevaArXiv2015} The detailed analysis of this phenomenon is, however, beyond the scope of the present work.

%%%%%%%%%%%%%%%%%%%%%%%%%%%%
%%%% fig:fig3
%%%%%%%%%%%%%%%%%%%%%%%%%%%%
\begin{figure}[t]
\centerline{\includegraphics[width=0.4\textwidth]{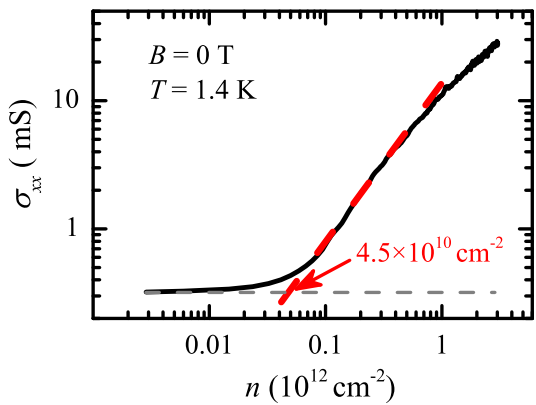}}
\caption{Longitudinal conductivity $\sigma_{xx}=1/\rho_{xx}$ as a function of charge carrier density $n$ for $T=1.4$\,K and $B=0$ for hole doping. The red dashed line is a fit to the expected linear behavior for $n \gg n_\textrm{Q}^*$, i.e.~$\log \sigma_{xx}   = \log n + const.$ The intercept of the linear fit with the value of the residual conductivity (horizontal gray dashed line) indicates the residual quasiparticle density $n_\textrm{Q}^*$ due to electrostatic potential fluctuations.}
\vspace{0em}
\label{fig:fig3}
\end{figure}
%%%%%%%%%%%%%%%%%%%%%%%%%%%%%

The transport properties at the CNP are governed by the quasi-particle density $n_\textrm{Q}$. In an actual device, when the gate voltage is swept across the charge neutrality region, the quasiparticle density saturates around a {non-zero} value $n_\textrm{Q}^*$ which is the minimum quasiparticle density that can be achieved experimentally. An estimate of $n_\textrm{Q}^*$ can be obtained from the measurement of the conductivity $\sigma_{xx}$ of the device in zero magnetic field\cite{DuNatNano2008}. In Fig.\ref{fig:fig3} we show $\sigma_{xx}$ as a function of the charge carried density $n$ at $0$\,T and $1.4$\,K. Around the CNP the conductivity saturates at the value $\sigma_{xx}=3.2 \times 10^{-4}$\,S, which is indicated by the horizontal dashed line. The intersection of this dashed lines with the linear fit to $\log(\sigma_{xx})$ provides us with an estimate for the minimal quasiparticle density  $n_\textrm{Q}^{*}= 4.5\times 10^{10}$\,cm$^{-2}$ at the CNP.

We find this value to be much larger than what is expected in the case of thermally excited quasiparticles in clean graphene. If we consider the density of state $\nu(\ep)$, from Eq.~\eqref{dens}  at finite temperature and zero magnetic field we obtain $n_\textrm{Q}(n\! =\! 0)= \pi T^2/3\hbar^2v^2$. For $T=1.4$\,K this amounts to $n_\textrm{Q}(n\! =\! 0)= 3.5\times 10^{6}$\,cm$^{-2}$. This value is four orders of magnitude smaller than $n_\textrm{Q}^{*}$, meaning that the realistic density of states $\nu(\ep)$ at the CNP is much larger than the one for ideal graphene and that the origin of the large quasiparticle density is intrinsic of the device. The most obvious reason for a finite {non-zero} value of the density of states in the vicinity of the Dirac point is the electrostatic potential variation induced e.\,g. by charged (or Coulomb) impurities.\cite{AdamPNAS2007}

The Zeeman effect provides a competing mechanism which induces a {non-zero} density of states at the CNP. For ideal graphene at zero temperature one finds from Eq.~\eqref{dens} that $n_\textrm{Q}(n\! =\! 0)=E_Z^2/4\pi \hbar^2v^2$. For a field of $30$\,T this estimate gives the figure $n_\textrm{Q}(n\! =\! 0)= 2.2\times 10^8$\,cm$^{-2}$ which is, however, still two orders of magnitude smaller than $n_\textrm{Q}^*$. Despite the low temperature and the large $B_\parallel$ employed in the experiment such that $T\ll E_Z$, the Zeeman splitting is most likely masked by the potential fluctuations around the CNP and therefore cannot be detected in our experiment.

Finally, we can also compare the energy broadening at the CNP responsible for the smearing out of the effects of Zeeman splitting in a parallel magnetic field with the Landau level broadening of the same sample in a perpendicular magnetic field estimated to be  $\Gamma = 14$~K.\cite{ChiappiniPRB2015} This is comparable to the expected spin splitting at 30 T. However, one should realize that the experiments to determine Landau level broadening are performed far away from the CNP where screening effects can significantly reduce potential fluctuations. Therefore our method of determining $n_\textrm{Q}^*$ at the CNP is more reliable. Indeed, using the ideal DOS of graphene and the residual carrier concentration one can estimate an energy smearing at the CNP which is more an order of magnitude larger than the one extracted from Landau level broadening.

Owing to the development in the device fabrication technique, it is nowadays possible to achieve the quasiparticle density in graphene to be as low as $10^8$\,cm$^{-2}$.\cite{CrossnoScience2016}  We may, therefore, expect that new experiments will soon be able to address the spin physics of graphene in a parallel magnetic field.

\section{Summary}

In conclusion, we have measured the resistivity of graphene on h-BN in a parallel magnetic field. At high charge carrier concentrations we do not observe any dependence of $\rho_{xx}$ on the external magnetic field and we demonstrated that all the changes observed at low $n$ and at the CNP can be ascribed to $B_\perp$. This indicates that the large parallel magnetic field up to $30$\,T and, consequently, Zeeman splitting up to $3.5$\,meV do not have any effect on the transport properties despite the rather high mobility $\mu \approx 50000$\,cm$^{2}$V$^{-1}$s$^{-1}$ in the sample. This observation is compatible with the leading role of Coulomb impurities in graphene that induce sizeable smooth variations of electrostatic potential at charge neutrality without reducing the mobility of charge carriers.\cite{KatsnelsonNatPhy2006}  We conclude that the presence of smooth electrostatic potential variation in the sample fully mask the effects of Zeeman splitting in our samples.

\begin{acknowledgments}
The was supported by the Dutch Science Foundation NWO/FOM 13PR3118 and by the EU Network FP7-PEOPLE-2013-IRSES Grant No 612624 ``InterNoM''. We acknowledge the support of the HFML-RU/FOM, member of the European Magnetic Field Laboratory (EMFL).
\end{acknowledgments}

\bibliography{Bibliography}

\end{document}